\documentclass{emulateapj}

\lefthead{van der Wel et al.}
\righthead{Structural Evolution of Early-Type Galaxies}
\slugcomment{The Astrophysical Journal, Accepted}

\begin{document}
\newcommand{\clll}{CL~1358+62~}
\newcommand{\msl}{MS~2053-04~}
\newcommand{\cll}{RX~J0152.7-1357~}
\newcommand{\ms}{MS~1054-0321~}
\newcommand{\lynx}{RDCS~0848+4453~}
\newcommand{\clh}{RDCS~1252-2927~}
\newcommand{\sige}{\sigma_{\rm{eff}}}
\newcommand{\reff}{R_{\rm{eff}}}
\newcommand{\kms}{\>{\rm km}\,{\rm s}^{-1}~}
\newcommand{\mvir}{M_{\rm{dyn}}}
\newcommand{\dens}{\Sigma_{\rm{eff}}}
\newcommand{\msol}{M_{\odot}}

\title{Recent Structural Evolution of Early-Type Galaxies: Size Growth
  from $z=1$ to $z=0$\altaffilmark{1}}

\author{Arjen van der Wel\altaffilmark{2}, Bradford
  P. Holden\altaffilmark{3}, Andrew W. Zirm\altaffilmark{2}, Marijn
  Franx\altaffilmark{4}, Alessandro Rettura{2}, Garth
  D. Illingworth\altaffilmark{3} \& Holland C. Ford\altaffilmark{2}}

\altaffiltext{1}{Based on observations with the \textit{Hubble Space
Telescope}, obtained at the Space Telescope Science Institute, which
is operated by AURA, Inc., under NASA contract NAS5-26555, and
observations made with the \textit{Spitzer Space Telescope}, which is
operated by the Jet Propulsion Laboratory, California Institute of
Technology, under NASA contract 1407. Based on observations collected
at the European Southern Observatory, Chile (169.A-0458). Some of the
data presented herein were obtained at the W.M. Keck Observatory,
which is operated as a scientific partnership among the California
Institute of Technology, the University of California and the National
Aeronautics and Space Administration. The Observatory was made
possible by the generous financial support of the W.M. Keck
Foundation. }

\altaffiltext{2}{Department of Physics and Astronomy, Johns Hopkins
 University, 3400 North Charles Street, Baltimore, MD 21218; e-mail:
 wel@pha.jhu.edu}

\altaffiltext{3}{University of California Observatories/Lick
  Observatory, University of California, Santa Cruz, CA 95064}

\altaffiltext{4}{Leiden Observatory, Leiden University, P.O.Box 9513,
  NL-2300 AA Leiden, Netherlands}

\begin{abstract}
  Strong size and internal density evolution of early-type galaxies
  between $z\sim 2$ and the present has been reported by several
  authors.  Here we analyze samples of nearby and distant ($z\sim 1$)
  galaxies with dynamically measured masses in order to confirm the
  previous, model-dependent results and constrain the uncertainties
  that may play a role.  Velocity dispersion ($\sigma$) measurements
  are taken from the literature for 50 morphologically selected
  $0.8<z<1.2$ field and cluster early-type galaxies with typical
  masses $\mvir = 2\times10^{11}~\msol$.  Sizes ($\reff$) are
  determined with Advanced Camera for Surveys imaging.  We compare the
  distant sample with a large sample of nearby ($0.04<z<0.08$)
  early-type galaxies extracted from the Sloan Digital Sky Survey for
  which we determine sizes, masses, and densities in a consistent
  manner, using simulations to quantify systematic differences between
  the size measurements of nearby and distant galaxies.  We find a
  highly significant difference between the $\sigma$-$\reff$
  distributions of the nearby and distant samples, regardless of
  sample selection effects.  The implied evolution in $\reff$ at fixed
  mass between $z=1$ and the present is a factor of $1.97 \pm 0.15$.
  This is in qualitative agreement with semianalytic models; however,
  the observed evolution is much faster than the predicted evolution.
  Our results reinforce and are quantitatively consistent with
  previous, photometric studies that found size evolution of up to a
  factor of 5 since $z\sim 2$.  A combination of structural evolution
  of individual galaxies through the accretion of companions and the
  continuous formation of early-type galaxies through increasingly
  gas-poor mergers is one plausible explanation of the observations.
\end{abstract}


\section{INTRODUCTION}\label{intro} 

Hierarchical galaxy formation models embedded in a $\Lambda$CDM
cosmology predict strong size evolution for massive galaxies.  A
higher gas fraction in high-redshift galaxies leads to more
dissipation and hence compact galaxies \citep[e.g.,][]{robertson06,
  khochfar06a}, and subsequent evolution such as dry merging or
accretion of smaller systems can increase the size of a galaxy
\citep[e.g.][]{loeb03, naab07}.  The models predict the strongest
sample-averaged size evolution for the most massive galaxies
\citep{khochfar06b} because of large differences in the gas fraction
at different redshifts and because the assembly of massive galaxies
continues until very late epochs in a hierarchical
framework\citep[e.g.,][]{delucia06}.

Evidence for significant size evolution between $z\sim 2$ and the
present has been building up quickly over the past few years
\citep[e.g.][]{trujillo04, trujillo06a, franx08}. In particular,
galaxies with low star formation rates and high stellar masses
($\gtrsim 10^{11}~\msol$) appear to be extremely compact from $z\sim
1.5$ \citep{daddi05, trujillo06b, trujillo07, longhetti07, cimatti08,
  rettura08} to $z\sim 2.5$ \citep{zirm07, toft07, vandokkum08b,
  buitrago08}.  Given the similarity between many of their observed
properties there is likely to be an evolutionary connection between
these distant compact galaxies and the present-day early-type galaxies
despite the measured large difference of 2 orders of magnitude in
surface mass density \citep[e.g.,][]{zirm07}.

The measurements of sizes and densities of high-redshift galaxies are
hampered by many systematic uncertainties (e.g., morphological
\textit{K}-corrections, surface brightness dimming, errors in
photometric redshifts, and mass measurements). Most of these errors,
however, are unlikely to fully account for the observed strong size
evolution.  The uncertainty in the mass estimates may be the
exception.  For the work in the literature, these mass estimates are
always based on the photometric properties of the galaxies.  For a
reasonable set of assumptions the photometric stellar mass estimates
are not uncertain by more than a factor of 2 or 3 and would not change
the inferred evolution significantly.  However, since we infer that
$z\sim 2$ galaxies must have physical central densities that are 3
orders of magnitude higher than those of local galaxies
\citep{vandokkum08b}, further verification of those apparently
reasonable assumptions is warranted.  For example, a stellar initial
mass function (IMF) that is radically different
\citep[e.g.,][]{larson05, fardal07, vandokkum08a, dave08} from a
Salpeter-like IMF \citep{salpeter55, scalo86, kroupa01, chabrier03,
  hoversten08} could reduce the stellar mass estimates by an order of
magnitude, producing perfectly normal galaxies by today's standards.

The spectacular nature of these compact galaxies at $z\sim 2$ could be
confirmed by direct, kinematical mass measurements.  However, the
quiescent nature of these objects and their consequent lack of
emission lines \citep{kriek06b} require absorption-line measurements
of their stellar velocity dispersions, which should be as high as
$400-500~\rm{km~s}^{-1}$ \citep{toft07, vandokkum08b}.  Unfortunately,
with the currently available instrumentation this is not
feasible. These $z \sim 2$ galaxies are prohibitively faint at
observed optical wavelengths \citep[see, e.g.,][]{cimatti08}, and
near-infrared spectroscopy is still maturing as a technique.
Continuum detections in the observed NIR have only recently become
possible \citep{kriek06b} for the brightest sources, and no detection
of absorption lines has been made.

At lower redshifts ($z\sim 1$) absorption-line spectroscopy has for
years been a powerful tool to study the evolution of distant
early-type galaxies \citep{vandokkum98, vandokkumstanford03,
  vandokkumellis03, wuyts04, vanderwel04, treu05a, holden05,
  vanderwel05, treu05b, diserego05, jorgensen05}. Size evolution is a
gradual process \citep[see, e.g.,][]{trujillo06a}; therefore,
intermediate changes in sizes and densities should be observable at
these redshifts.

In this paper we compile a sample of galaxies at redshifts $0.8<z<1.2$
with measured absorption line velocity dispersions from the literature
and that are visually classified as early-type galaxies with the aid
of \textit{Hubble Space Telescope} ({\it HST}) imaging from the
Advanced Camera for Surveys \citep[ACS;][]{ford98}. We measure the
galaxies' sizes from these ACS data.  We then compare this distant
sample with nearby early-type galaxies extracted from the Sloan
Digital Sky Survey \citep[SDSS;][]{york00}.  This comparison, with
careful control of systematic uncertainties, allows us to verify that
distant early-type galaxies are indeed significantly more compact than
their local counterparts.

The advantage of this approach is that the size and density
measurements are independent of the photometric properties of the
galaxies apart from the surface brightness profile.  The absence of
luminosity and other photometric properties from our analysis assures
us that our study does not suffer from the strong possible biases in
previous photometric work.  Moreover, deep, high-resolution ACS
imaging allows us to determine sizes of $z\sim 1$ galaxies to a
precision comparable to that nearby galaxies and in a consistent
manner.  Most previous studies verify for biases in the size
determinations within their distant samples
\citep[e.g.,][]{trujillo04, trujillo06a, cimatti08} but do not extend
this analysis to verify the consistency with size measurements of
nearby galaxies.

In \S~\ref{nearby} we describe the sample of nearby early-type
galaxies and derive the dynamical mass-size relation.  In
\S~\ref{distant} we construct the sample of $z\sim 1$ early-type
galaxies, determining their masses and sizes in a manner that is
consistent with the nearby sample.  In \S~\ref{sim} we quantify
systematic effects in our size measurements through simulations.  In
\S~\ref{results} we derive the evolution in the dynamical mass-size
relation.  In \S~\ref{dis} we compare our results with previous
measurements based on photometric mass estimates and semianalytic
model predictions, and we discuss size evolution in the broader
context of the evolving early-type galaxy population.  Finally, in
\S~\ref{sum} we summarize our results and conclusions.  We adopt the
following cosmological parameters:
$(\Omega_M,~\Omega_{\Lambda},~h) = (0.3,~0.7,~0.7)$.


\section{NEARBY EARLY-TYPE GALAXIES}\label{nearby}

\subsection{Velocity Dispersions and Sizes}\label{nearby:sigrad}

We have extracted a large sample of early-type galaxies at redshifts
$0.04<z<0.08$ from the SDSS database \citep[DR6;][]{adelman08} based
on the criteria as outlined by \citet{graves07}.\footnote{The IDs,
  positions, redshifts, and dispersions for this sample were kindly
  provided by G. Graves} Briefly, galaxies on the red sequence and
either without emission lines or with high
$[\rm{OII}]$~to~$\rm{H}\alpha$ ratios are included in the sample.
These criteria effectively exclude star-forming galaxies, but include
genuine early-type galaxies with nuclear activity
\citep[see][]{yan06}.

The dispersion as measured within the spectroscopic aperture
($\sigma_{\rm{ap}}$) is corrected to match the average dispersion
within the effective radius $\reff$ (measured as described below)
following \citet{jorgensen95spec}:
\begin{equation}
  \sige = \sigma_{\rm{ap}} \left(\frac{\reff}{R_{\rm{ap}}(z)}
  \right)^{-0.04},
\label{eq:apcor}
\end{equation}
where $R_{\rm{ap}}(z)$ is the radius of the SDSS spectroscopic fiber
($1.5''$) in kpc at the distance of the galaxy.  We use the correction
from \citet{jorgensen95spec} for consistency with previously published
results.  We note that \citet{cappellari06} used better data to
improve the aperture correction, but the resulting difference in
$\sige$ is only a few percent.

We use GALFIT \citep{peng02} to determine effective radii from
the SDSS $g$-band imaging assuming an $R^{1/4}$ profile, leaving the
effective radius, the integrated magnitude, the position angle, the
axial ratio, and the position of the center as free parameters. The
point-spread function (PSF), which is used to deconvolve the image, is
constructed for each galaxy separately by co-adding the stars in the
frames after drizzling the cutouts to a common center. A more general
$R^{1/n}$ profile \citep{sersic68} may provide a more realistic
description of the surface brightness distribution of individual
early-type galaxies, especially in the presence of a significant
disk. However, $n=4$ provides a good description of the average
profile of early-type galaxies both nearby \citep{devaucouleurs48} and
at $z\sim 1$ (see \S~\ref{distant:profile}).  Moreover, introducing
$n$ as an additional free parameter results in unnecessarily large,
redshift-dependent systematic uncertainties in the size measurements
(see \S~\ref{sim}).

The size parameter that we use in this paper is the circularized
effective radius $\sqrt{ab} \equiv a \sqrt{q}$, where $a$ is the
effective radius along the major axis (the output parameter of GALFIT,
$b$ is the effective radius along the minor axis, and $q$ the axis
ratio (as calculated by GALFIT); $\sqrt{ab}$ is a good approximation
for optically thin luminosity distributions such as the generally
dust-poor early-type galaxies in our samples.  The systematic and
random errors of our size determinations are inferred from extensive
simulations described in \S~\ref{sim}.

The SDSS spectroscopic catalog suffers from several biases that may
mitigate size evolution measurements.  First, compact sources are not
targeted for spectroscopy as they may be mistaken for stars or because
their central surface brightnesses, i.e., their fiber magnitudes, are
too bright.  Second, almost all galaxies that in the literature
\citep[see the HyperLEDA database compiled by][]{paturel03} have been
claimed to have high, $>300\kms$ velocity dispersions have dispersions
of $<300\kms$ in the SDSS \citep[see][Appendix A]{bernardi07a}.  The
source of this discrepancy is unknown.  While Bernardi convincingly
argues that the SDSS dispersions are more reliable, there are a number
of galaxies with large, mutually consistent dispersion measurements
from multiple, independent observers, and for which the SDSS
dispersion is significantly smaller.  These potential caveats may
cause our size evolution measurements to be biased.  We refer to these
issues when we present our results in \S~\ref{results}.

\subsection{The Mass-Radius Relation and the Mass-Density Relation}\label{nearby:mr}

From $\sige$ and $\reff$ we derive the total dynamical mass and the
corresponding average surface mass density within $\reff$:
\begin{equation}
  \mvir = \frac{\beta \reff \sige^2}{G} ,
\label{eq:mass}
\end{equation}
\begin{equation}
  \dens = \frac{\beta \sige^2}{2 \pi G \reff} ,
\label{eq:dens}
\end{equation}
with $\beta=5$, which has been shown to hold for local galaxies
\citep{cappellari06}.  Following \citet{shen03} we adopt the following
characterization of the $\mvir$-$\reff$ relation:
\begin{equation} 
  \reff=R_c\left(\frac{\mvir}{M_c}\right)^b.
\label{eq:mr}
\end{equation}
With a least-squares linear fit to all galaxies with mass $\mvir >
3\times 10^{10}~\msol$ we find that the slope is $b=0.56$ and the zero
point normalized to a characteristic mass $M_c=2\times 10^{11}~\msol$
is $R_c=4.80$~kpc.  We find statistically the same relation if we
perform a linear fit to the values of the median $\reff$ in 0.1 dex
wide mass bins in the range $10.5 < \log(\mvir) < 12.1$.  The scatter
around the best-fit relation is 0.14 dex.

Using stellar masses, $M_*$, \citet{shen03} find the same slope
$b=0.56$ for the $M_*$-$\reff$ relation.  Their zeropoint, however, is
larger ($R_c=6.14$~kpc).  This is likely due to the difference between
$\mvir$ and $M_*$ as \citet{cappellari06} show for a \citet{kroupa01}
IMF (which is also used by Shen et al.~2003) that $\mvir\sim 1.4M_*$.
This translates into a difference in $R_c$ of $\sim$20\%, close to the
observed difference.  Furthermore, \citet{shen03} analyze SDSS
$r$-band imaging whereas we use $g$-band imaging.

\section{DISTANT EARLY-TYPE GALAXIES}\label{distant}

\subsection{Velocity Dispersions and Sizes}\label{distant:sigrad}

Several authors have published velocity dispersion measurements of
early-type galaxies at $z\sim 1$.  We compile the data from three
different data sets for which the selection criteria are well
understood so that systematic effects introduced through selection
effects can be properly modeled.  Our compiled sample contains
galaxies in the redshift range $0.8<z<1.2$ in the Chandra Deep
Field-South \citep[CDF-S,][]{vanderwel04, vanderwel05} and the Hubble
Deep Field-North \citep[HDF-N,][]{treu05a, treu05b}. In addition, we
include galaxies in the massive, X-ray selected cluster \ms at
$z=0.831$ \citep{wuyts04}.  The seven cluster galaxies at $z>1$ for
which dispersions have been measured \citep{vandokkumstanford03,
  holden05} are not included because of the paucity of this sample,
which prevents us to accurately model selection effects.  The final
sample only contains galaxies with $S/N>10~\rm{\AA}^{-1}$ since
dispersions derived from spectra with lesser quality can suffer from
large ($>10\%$) systematic uncertainties.  The same aperture
corrections are included as for the nearby galaxies
(Eq.~\ref{eq:apcor}), with in this case $R_{\rm{ap}}(z)$ the radius of
a circle with area $1''\times 1.25''$ (the width of the slits and the
typical length of the extracted region) in kiloparsec at the redshift
of the galaxy. The data are given in Table \ref{tab}.

For all galaxies ACS imaging is available. In order to produce an
internally consistent data set, we re-measure the sizes of all
galaxies.  GOODS\footnote{See http://www.stsci.edu/science/goods/}
provides deep, publicly available ACS imaging of the CDF-S and the
HDF-N \citep{giavalisco04} in four filters. We use the F850LP
($z_{850}$-band) images in order to match the rest-frame wavelength at
which the sizes of the nearby comparison sample are measured (the SDSS
$g$-band; see \S~\ref{nearby}).  For the \ms cluster ACS imaging has
been taken as part of the guaranteed time observation program
\citep{blakeslee06}.  We use the F775W ($i_{775}$-band) imaging as the
available $z$-band imaging is of lesser quality.  At this redshift
rest frame $g$ falls in between $i_{775}$ and $z_{850}$ such that the
morphological \textit{K}-correction is not a problem; the $z\sim 0.8$
galaxies in the sample of \citet{treu05b} are only $3\% \pm 4\%$
smaller in the $i_{775}$ band than in the $z_{850}$ band.

With GALFIT we determine effective radii in the same manner as for the
nearby galaxy sample (see \S~\ref{nearby:sigrad}). The PSF is
constructed with Tiny Tim \citep{krist95}, even though using stars
results in negligible differences \citep[see,~e.g.,][]{vanderwel05,
  treu05b}.  Errors are discussed in \S~\ref{sim}, and the data are
given in Table \ref{tab}.

We use Eqs.~\ref{eq:mass} and \ref{eq:dens} to compute masses and
surface densities. Again, we adopt $\beta=5$, which has been shown to
hold for distant nonrotating elliptical galaxies
\citep{vandermarel07b, vanderwel08b}.  For rotating early-type
galaxies the situation appears to be more complex \citep{vanderwel08b}
in the sense that $\beta$ is possibly $\sim$20\% larger than 5.  We
comment on the impact of this possible complication on our size
evolution measurement in \S~\ref{results:mrevol}.  A low-mass cutoff
of $3\times 10^{10}~\msol$ is applied since below this limit no useful
samples are available due to severe incompleteness of the surveys
\citep[see, e.g.,][]{vanderwel05}.

\subsection{The Average Surface Brightness Profile}\label{distant:profile}

In determining the sizes of the nearby and distant galaxies in the
previous sections we assumed that an $R^{1/4}$ profile provides an
accurate description of early-type galaxies.  We know this to be true
for nearby galaxies, but not for $z\sim 1$ early-type galaxies.
If a more general $R^{1/n}$ profile is adopted, measured values tend
to cluster around $n=4$ \citep[see, e.g.,][]{blakeslee06, rettura06}.
However, there is a possibility that the true values of $n$ are
different; at large radii the ``wings'' of the profile become quickly
overwhelmed by background noise, even in the deepest \textit{HST}
imaging.  Because $n$ and the measured $\reff$ are correlated,
assuming $n=4$ for all redshifts introduces systematic errors in case
$n$ evolves with redshift.

\begin{figure}[t]
\epsscale{1.1} 
\plotone{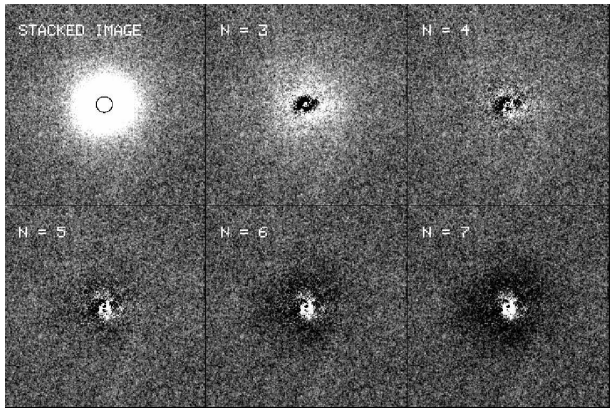}
\caption{ Stacked $z_{850}$-band image of 29 $z\sim 1$ elliptical
  galaxies in the CDF-S and the HDF-N and the residuals after
  subtracting $R^{1/n}$ model profiles, with $n=$ 3, 4, 5, 6, 7.  Each
  panel is 7.68" on a side, which corresponds to 62 kpc at $z=1$.  The
  circle indicates the effective radius as measured with the $R^{1/4}$
  model profile (0.31", or 2.5 kpc at $z=1$).  The model with $n=4$
  provides the best fit.  The model with $n=3$ produces positive
  residuals at large radii; models with high $n$ produce negative
  residuals.  This justifies our choice to adopt the $R^{1/4}$ law to
  model the surface brightness profiles of both local and distant
  early-type galaxies. }
\label{stack}
\end{figure}

In order to examine the profiles of the $z\sim 1$ galaxies at large
radii, we median-stack the $z_{850}$-band images of all elliptical
galaxies (S0s are excluded) without bright neighbors in our CDF-S and
HDF-N samples (see Fig.~\ref{stack}).  The images of the individual
galaxies are drizzled onto a common central position. Due to
imperfections in this procedure, the stacked PSF may not be an
accurate description of the PSF of the stacked image.  However, this
does not play a role since the deviations we are interested in have
scales that are an order of magnitude larger than the PSF.

With GALFIT we subtract $R^{1/n}$ profiles with integer values $n=3-7$
(see Fig.~\ref{stack}).  The negative residuals outside $\reff$~for
models with large $n$ and the positive residuals for models with small
$n$ indicate that these limiting cases provide poor fits of the outer
regions of elliptical galaxies at $z\sim 1$.  The $R^{1/4}$ and
$R^{1/5}$ profiles provide the best description of their average
surface brightness distributions. This visual impression is confirmed
by the $\chi^2$-values of the respective fits: $\chi^2=0.5$ for both
$n=4$ and for $n=5$, whereas $\chi^2>0.7$ for other values of $n$.
Interestingly, $n$ does not evolve significantly with redshift, and we
conclude that it is safe to assume that choosing $n=4$ for both nearby
and distant early-type galaxies does not introduce significant
systematic errors.

\section{SIMULATIONS OF SIZE MEASUREMENTS}\label{sim}

To test the robustness of our size determinations of local and distant
early-type galaxies in \S\S~\ref{nearby} and \ref{distant} we simulate
size measurements by using SDSS $g$-band imaging of 45 early-type
galaxies in the Virgo Cluster \citep{mei07}.  The pixels of the
mosaics of the Virgo Cluster galaxies are re-binned to account for the
different pixel scales of the various instruments, and different
cosmological distances of the galaxies at higher redshifts. The
redshift range is $z=0.04-0.08$ for the nearby sample and $0.8-1.2$
for the distant sample. The physical sizes of the simulated galaxies
are thus conserved.  For each redshift ($z=0.04, 0.06, 0.08, 0.8,
1.2$) we run $\sim$200 simulations with different values for the flux
density of the simulated galaxies, which are chosen such that the
simulated galaxies have the same range in apparent magnitude as the
observed galaxies in our samples.  After convolution with the
appropriate PSF and the addition of Poisson noise the seed galaxies
are inserted into empty parts of real images.  Their sizes are
measured with GALFIT in the same manner, by fitting a $R^{1/4}$ law,
as for the real galaxies.  Because we are mainly interested in the
systematic differences in the size determinations within and between
our nearby and distant samples, we assume the size determinations
based on the $z=0.04$ simulated SDSS images of the Virgo Cluster
galaxies as the baseline against which we compare the other simulated
size measurements.

\begin{figure}[t]
\epsscale{1.2} 
\plotone{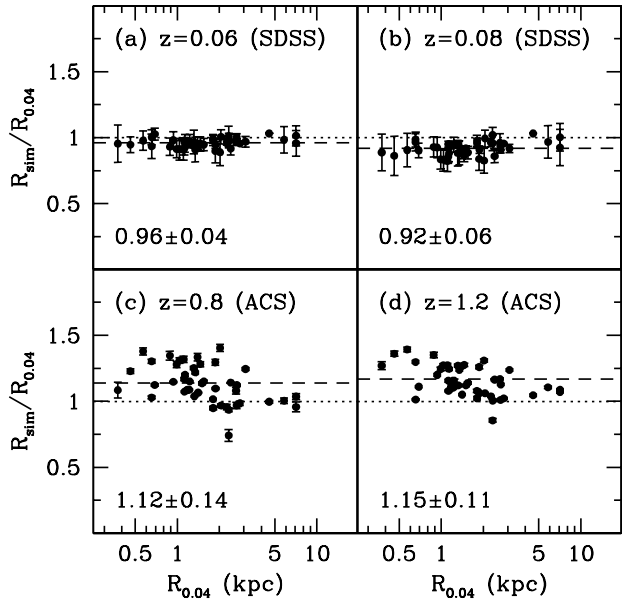}
\caption{ Random and systematic errors in the size determinations of
  early-type galaxies with SDSS imaging at ({\it a}) $z=0.06$ and
  ({\it b}) $z=0.08$ and with \textit{HST} imaging at ({\it c})
  $z=0.8$ and ({\it d}) $z=1.2$, all with respect to the size
  measurements at $=0.04$, which are used as the benchmark in our
  analysis. These are the results of simulations with 45 early-type
  galaxies in the Virgo Cluster. The systematic offsets are indicated
  by the dashed lines. The scatter in the offsets, considered to be
  the random error in the size determinations, are also listed.}
\label{simul}
\end{figure}

\begin{figure*}[t]
\epsscale{1.1}
\plottwo{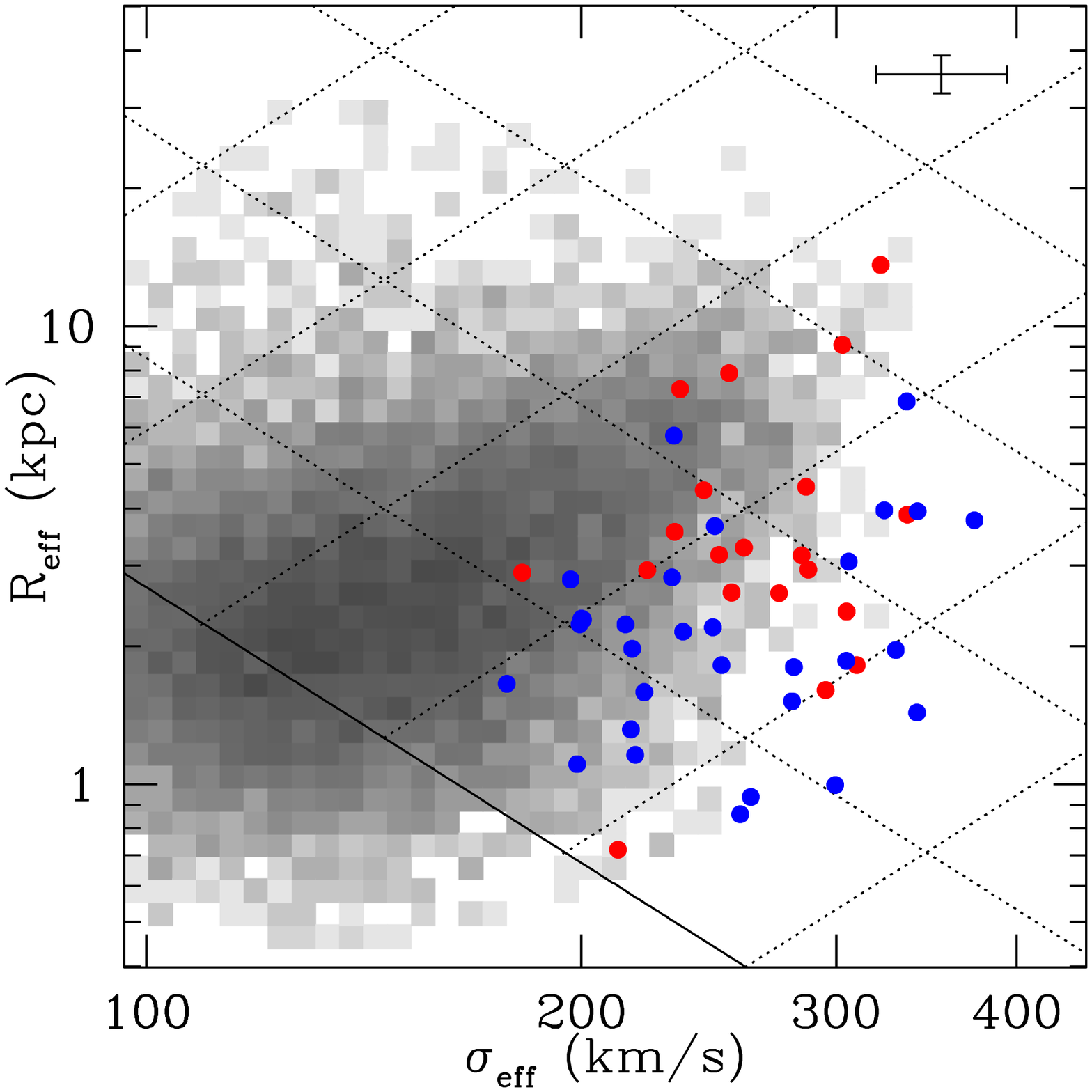}{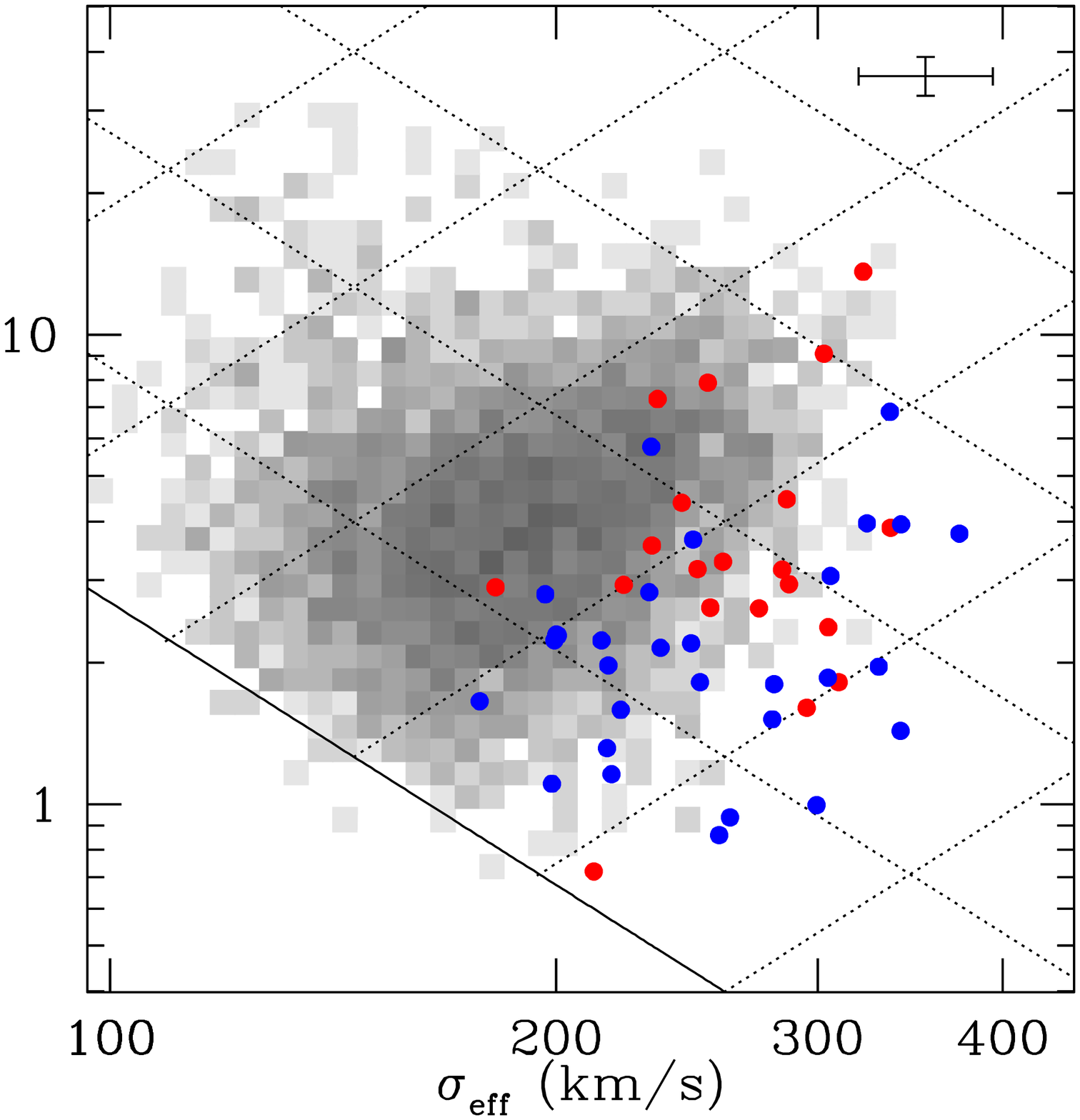}
\caption{ The $\sige$-$\reff$ distributions of the nearby sample of
  early-type galaxies (\textit{gray scale}) and the distant sample
  (data points). The red data points are cluster galaxies, and the
  blue data points are field galaxies.  The error bars at the top
  right indicate the typical values of the errors for the distant
  sample.  The solid line indicates $\log(\mvir)=10.5$.  Dotted lines
  indicate lines of constant $\mvir$ (\textit{parallel to the solid
    line}), spaced by 0.5 dex, and lines of constant surface density
  $\dens$ (\textit{perpendicular to the solid line}), also spaced by
  0.5 dex.  The left-hand panel shows the entire nearby sample; the
  right-hand panel only shows those galaxies in the nearby sample that
  would be included in the distant sample considering the selection
  effects, that apply to the surveys (see \S~\ref{results:sigrad}).
  The highly significant offset ($>99.9\%$ significance) between the
  local and distant samples implies significant structural evolution
  in the early-type galaxy population between $z\sim 1$ and the
  present. }
\label{sig_rad}
\end{figure*}

The results of the simulations are shown in Fig.~\ref{simul}.  Within
the nearby sample we find systematic, redshift-dependent differences,
of up to $\sim$10\%.  Random errors, derived from the scatter in the
sizes inferred from the simulated images, are typically less than 5\%.
Systematic differences between the nearby and distant samples can be
as large as 20\% for small galaxies, where at high redshift the sizes
are overestimated.  Random errors are typically 10--15\%. We find no
systematic trends with magnitude. The reason for this is that all
galaxies are relatively bright compared to the depth of the data sets,
such that the limiting factor in the size measurements is spatial
resolution.

Adopting a $R^{1/n}$ law with $n$ as a free parameter may improve the
quality of the fits to individual galaxies.  However, our simulations
reveal that the random errors increase to $\sim$20--25\%, without much
change in the systematic errors.  Together with the analysis of
stacked images (\S~\ref{distant:profile}), this test justifies our
choice to use the $R^{1/4}$ law to describe the surface brightness
profiles of all galaxies in both the nearby and the distant samples.

The sizes we derive in the \S\S~\ref{nearby} and \ref{distant}, and
the derived quantities $\mvir$ (Eq.~\ref{eq:mass}) and $\dens$
(Eq.~\ref{eq:dens}), are corrected to account for systematic
measurement errors.  Those corrections depend on redshift and are
interpolated between the values listed in Fig.~\ref{simul}.  For
simplicity the dependence on size is not taken into account, such that
the remaining systematic uncertainty is $\sim$5\%.

\section{STRUCTURAL EVOLUTION OF EARLY-TYPE GALAXIES}\label{results}

\subsection{Evolution of the $\sigma$-Radius Distribution}\label{results:sigrad}

In Figure \ref{sig_rad} we compare the $\sige$-$\reff$ distributions
of the nearby and distant early-type galaxy samples.  This unusual
projection of the fundamental plane \citep[FP;][]{dressler87,
  djorgovskidavis87} has a very large scatter.  However, the advantage
is that changes with redshift are independent of luminosity evolution.
Despite the large scatter, it is clear that the distant galaxies are
offset from the nearby galaxies.  Galaxies with the properties of
typical galaxies in the distant sample ($\sige\sim 250~\kms$;
$\reff\sim 3$ kpc) are rare in the local universe.  In the nearby
sample, galaxies with $\sige\sim 250~\kms$ have much larger sizes, and
galaxies with sizes $\reff\sim 3$ kpc have dispersions of $\sige\sim
150~\kms$. These numbers are only intended to guide the eye. A
quantitative analysis of the differences between nearby and distant
galaxies is presented below.

The distant sample is not directly comparable with the nearby sample
in its entirety (\textit{left}), as the nearby sample reaches to much
lower masses.  In order to assess the question whether the true,
underlying $\sige$-$\reff$ distribution of distant galaxies is
different from the $\sige$-$\reff$ distribution of nearby galaxies, we
need to remove the galaxies in the nearby sample that would not be
included at $z\sim 1$ due to sample selection effects.  The sub-sample
of nearby galaxies that is observable at $z=1$ is shown in the right
panel of Fig.~\ref{sig_rad}.  The two criteria that the galaxies in
the observable sub-sample satisfy are $L>L_{\rm{min}}$ and $\reff <
R_{\rm{eff,max}}$.  $L_{\rm{min}} \sim 10^{10} L_{\odot,\rm{B}}$ is
the luminosity limit for the field $z\sim 1$ surveys \citep[see,
e.g.,][]{vanderwel05} after correcting for luminosity evolution
between $z=1$ and the present \citep[0.555 dex;][]{vandokkum07}.  For
the \ms cluster sample from \citet{wuyts04} this is $1.8\times
10^{10}L_{\odot,\rm{B}}$.  The second criterion $\reff <
R_{\rm{eff,max}}$ takes into account that high signal-to-noise ratio
($S/N$) spectra are harder to obtain for low surface brightness
galaxies than for high surface brightness galaxies with the same
luminosity; i.e., the distant sample is biased in favor of small
galaxies.  The $S/N$ of the spectra of \citet{vanderwel05} and
\citet{treu05b} do not precisely scale linearly with luminosity
$L=I\reff^2$, where $I$ is the surface brightness, but as $S/N \propto
I\reff^{1.6}$. This implies that, at fixed luminosity $L$, $S/N
\propto R^{-0.4}$. Since a dispersion measurement requires a minimum
$S/N$ ($\sim 12 \AA^{-1}$), a galaxy with luminosity $L$ has a maximum
radius $R_{\rm{eff,max}} \propto L^{2.5}$ for which its dispersion can
be determined. We use the luminosity limits of the surveys discussed
above to normalize the dependence between luminosity and maximum size;
we simply assume that for the smallest galaxies ($\reff = 1$~kpc) the
luminosity limit coincides with the size limit such that we have
\begin{equation}
  R_{\rm{eff,max}}~(\rm{kpc})=\left(\frac{L}{L_{\rm{min}}}\right)^{2.5}.
\label{eq:sel}
\end{equation}

One would expect that for galaxies smaller than 3 kpc the
signal-to-noise ratio of the spectra would not depend on size any
longer since seeing generally dominates the apparent sizes of such
small galaxies at $z\sim 1$.  Because of the variety of telescopes,
weather conditions, and data reduction techniques, this, however, is
washed out and not apparent in the data.  We note that the
introduction of, effectively, a rudimentary surface brightness
criterion is a step forward in modeling the selection effects with
respect to earlier attempts that only take total luminosity into
account.

The difference between the $\sige$-$\reff$ distributions at low and
high redshift is highly significant, even after taking selection
effects into account (Fig.~\ref{sig_rad}, \textit{right}).  The
two-dimensional Kolgomorov-Smirnov statistic has a high value
($D=3.71$), which implies that it is extremely unlikely that the nearby
and distant samples are drawn from the same distribution.  By
repeatedly drawing samples from the nearby sample with the same size
as the distant sample we confirm this: less than 0.001\% of the
simulated samples have $D=3.71$ or higher.

\begin{figure}[t]
\epsscale{1.2}
\plotone{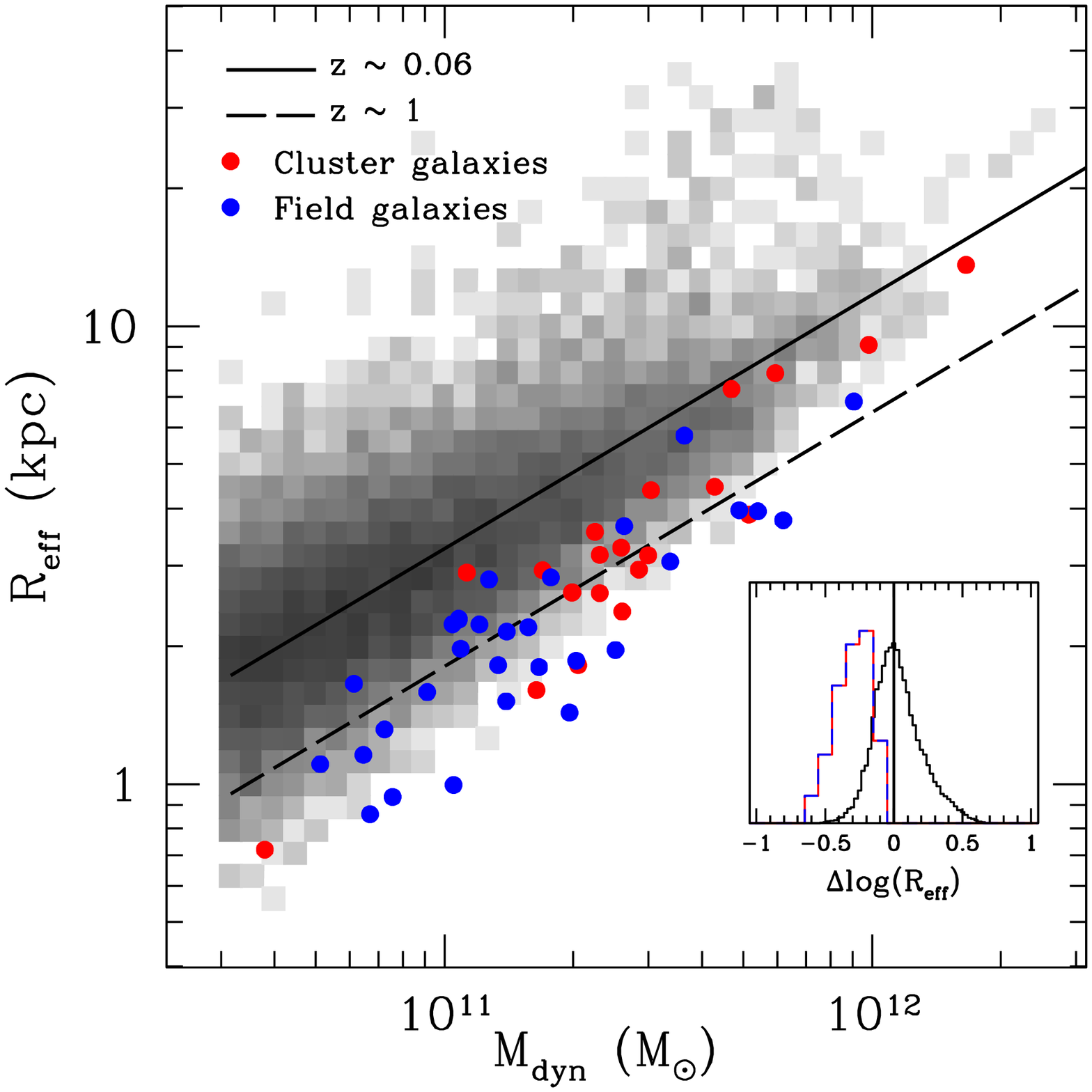}
\caption{Mass-size relation for the nearby sample (\textit{solid
    line}) and at $z\sim 1$ (\textit{dashed line}); the symbols are
  the same as in Fig.~\ref{sig_rad}.  For the derivation of the
  $\mvir$-$\reff$ relation for the nearby sample see
  \S~\ref{nearby:mr}; for the derivation of the $\mvir$-$\reff$
  relation for the distant sample see \S~\ref{results:mrevol}.  The
  smaller, inset panel shows the distribution of the two samples
  around the $\mvir$-$\reff$ relation of the nearby sample
  (\textit{the solid line in the large panel}).  The distant galaxies
  are $1.8\pm0.1$ times smaller than the nearby galaxies. It appears
  that the most massive galaxies do not show as large an offset. This
  indicates that size evolution may be slower for the highest mass
  galaxies than for low-mass galaxies, but it has to be kept in mind
  that these very massive galaxies are brightest cluster galaxies and
  may therefore have developed differently from other galaxies.}
\label{M_rad}
\end{figure}

\subsection{Evolution of the Mass-Radius Relation}\label{results:mrevol}

The structural difference between the nearby and distant samples
described in the previous section implies that the $\mvir$-$\reff$ and
$\mvir$-$\dens$ relations evolve with redshift.  In Fig.~\ref{M_rad}
we show the $\mvir$-$\reff$ relation for the distant sample, and
compare this with the equivalent relation for nearby galaxies derived
in \S~\ref{nearby:mr}.  Clearly, the relation shifts to smaller
radii from low to high redshift.

Parametrized as in Eq.~\ref{eq:mr} we find $R_c=2.58\pm0.17$
and $b=0.65\pm0.06$ with a scatter of $0.117\pm0.013$ dex (after
subtracting the observational uncertainties in quadrature).  The
errors are estimated with a bootstrap/Monte-Carlo simulation in which
the data points are randomly sampled and varied according to the
(correlated) measurement errors which are assumed to be Gaussian.  The
systematic error in $\reff$ of 5\% (see \S~\ref{sim}) is also taken
into account.

The treatment of the selection effects described in the previous
section shows that the observed size evolution seen in
Fig.~\ref{M_rad} is not an artifact.  However, given the nature of the
selection effects, which favor small galaxies over large galaxies, the
intrinsic amount of size evolution and possible evolution in the slope
and scatter of the $\mvir$-$\reff$ relation must be inferred through
careful modeling.  The goal is to derive the intrinsic $\mvir$-$\reff$
relation at $z\sim 1$ that reproduces the observed $\mvir$-$\reff$
distribution after applying the selection criteria.  We take an
iterative approach due to the interdependence of the selection
criteria and the amount of evolution in zeropoint, slope, and scatter
of the $\mvir$-$\reff$ relation.  In the following we de-evolve the
properties of the nearby sample to constrain the form of the true,
underlying $z\sim 1$~$\mvir$-$\reff$ relation.

The simplest evolutionary scenario is a change in the zero point $R_c$
(see Eq.~\ref{eq:mr}).  For each object in the nearby sample the size
is reduced by the same amount $\Delta\log(\reff)$, and those that do
not satisfy the selection criteria described in the previous section
are removed. From the remaining sub-sample the ``observed''
$\mvir$-$\reff$ relation is determined.  The different selection
criteria and sample sizes for field and cluster galaxies are taken
into account in this process, which is repeated for many different
values of $\Delta\log(\reff)$.  We find that an intrinsic value of
$R_c=2.64\pm0.18$ reproduces the observed value of $R_c=2.58\pm0.17$.
Hence, it appears that selection effects do not strongly affect the
inferred size evolution.

However, the scatter of the assumed intrinsic distribution (0.14 dex)
is higher than the observed scatter ($0.117\pm0.013$ dex).  This
cannot be explained by selection effects in the simple scenario
described above.  It is therefore required that the scatter, as well
as the zero point, is also treated as an evolving parameter.  This is
implemented in our analysis by reducing or increasing the offset of
each galaxy in the nearby sample from the best-fit $\mvir$-$\reff$
relation by a given fraction.  Doing so, we find that the best-fitting
zero point $R_c$ is not different from the earlier estimate based on a
nonevolving scatter.  We also find that the evidence for evolution in
the scatter is weak ($\sim 1.5~\sigma$).  This exercise mainly serves
to show that our size-evolution result in not sensitive to the amount
of evolution in the scatter allowed by the observations.

A similar verification must be carried out for evolution in the slope
of the $\mvir$-$\reff$ relation.  Allowing only the scatter and the
size to evolve, as described above, the inferred slope of the
``observed'' $\mvir$-$\reff$ relation is 0.59, marginally consistent
with the true observed slope of $b=0.65\pm0.06$.  If we treat the
slope as an additional, third free parameter we confirm that evolution
in the slope, as constrained by our measurements, does not affect our
size-evolution measurement.  An intrinsic slope of $b=0.61$ provides a
better fit than the original slope of the $\mvir$-$\reff$ relation of
the nearby sample ($b=0.56$), but the difference is marginal ($\sim
1~\sigma$).

We conclude that, despite (weak) evidence for evolution in the slope
and the scatter of the $\mvir$-$\reff$ relation with redshift, there
is no significant improvement in modeling the observations by adopting
slope and scatter as free parameters.  Modeling the evolution by a
fractional change in size, regardless of mass and offset from the
local $\mvir$-$\reff$ relation, provides an equally good fit.  Most
importantly, changing the slope and scatter within the range allowed
by the observations does not affect the inferred size evolution.  We
find that $R_c=2.64\pm0.18$ kpc at $z=0.90$, a factor of
$1.8\pm 0.1$ times smaller than at $z\sim 0.06$.

The weak evidence for a change in slope of the $\mvir$-$\reff$
relation may also be interpreted as a difference between field and
cluster galaxies, as the more massive galaxies in our sample tend to
be cluster galaxies.  Assuming that slope and scatter remain constant
but that the zero point of the $\mvir$-$\reff$ relation evolves
differently for field and cluster galaxies, we find that $R_c=2.49$
kpc for field galaxies and $R_c=3.06$ kpc for cluster galaxies.  The
1~$\sigma$ error on this difference of 0.57 kpc is 0.32 kpc.  The true
error may be larger since in this estimate it is assumed that scatter
and slope behave the same in the different environments and that there
are no relative systematic errors in the size determinations of field
and cluster galaxies.  The evidence for a difference between the size
evolution of field and cluster early-type galaxies is therefore weak
\citep[see also][]{rettura08}. However, we have to keep in mind that
so far only a very small number of clusters is considered. Future
studies will need to extend the existing analyses to a larger number
of clusters to verify the general validity of the results.

So far, we have assumed that the masses of the galaxies do not change.
Our justification is that the scatter hardly depends on mass; the
effect of a changing mass function on modeling selection effects is
expected to be small.  However, physically speaking, it is unnatural
to propose size evolution without changes in the masses of galaxies.
Moreover, if the characteristic mass above which the number density of
galaxies drops off exponentially evolves with redshift, selection
effects will change as well.  The simplest way to implement mass
evolution is to assume that $M\propto \reff$ (Eq.~\ref{eq:mass}).
Including this in our modeling procedure shows that the effect on the
inferred size evolution is less than 5\%, and we therefore adopt the
results with no mass evolution.

We recall that the nearby sample is biased against compact early-type
galaxies (\S~\ref{nearby:sigrad}).  The potentially underestimated
number of galaxies with dispersions $\sigma>300~\kms$ is unlikely to
drastically affect the size-evolution determination for the sample as
a whole as the average dispersion of the galaxies in the distant
sample is smaller than that.  However, the slope of the local
$\mvir$-$\reff$ relation is possibly overestimated, which would lead
to an underestimate of the slope evolution.  More important is the
problem that small galaxies are missed because of their photometric
misclassification as stars in the SDSS.  To fully address this issue a
complete analysis of the SDSS photometric catalog is required, which
is clearly beyond the scope of this paper.  However, we can say that
it is highly unlikely that the average size of nearby early-type
galaxies is underestimated by a factor of 2 because of this bias.
On the other hand, for the interpretation of our results and
identifying the mechanisms responsible for size evolution (see
\S~\ref{dis:3}) this bias could prove to be important.

In \S~\ref{distant} we noted that the dynamical mass estimate as
adopted in this paper (Eq.~\ref{eq:mass}, with $\beta=5$) may be too
low for rotating early-type galaxies.  If this is the case, then size
evolution for these galaxies will be underestimated by $\sim$10\%.
Since this is within the uncertainties of our measurements we do not
take this further into account.

\begin{figure}[t]
\epsscale{1.2}
\plotone{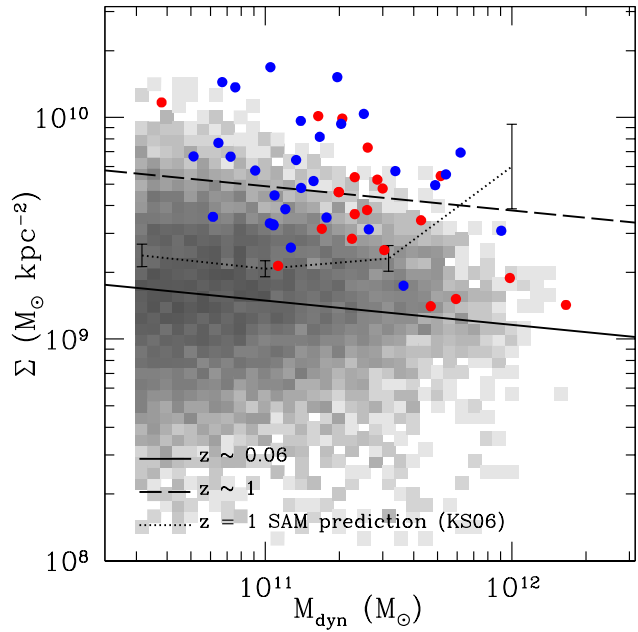}
\caption{Mass-density relation at $z\sim 1$. The symbols and lines are
  the same as in Figs.~\ref{sig_rad} and \ref{M_rad}.  The $z\sim 1$
  early-type galaxies are $\sim 4$ times more dense than their nearby
  counterparts. The prediction of the semianalytic size-evolution
  model for elliptical galaxies from \citet{khochfar06b} is shown as
  the dotted line.  The error bars indicate the predicted size
  evolution between $z=0.8$ and $z=1.2$, the redshift range of our
  distant sample.  Despite qualitative agreement, there are
  significant quantitative differences between the predicted and
  observed evolution.}
\label{M_dens}
\end{figure}

Obviously, size evolution at fixed mass translates into density
evolution. This is illustrated in Fig.~\ref{M_dens} where we compare
the density distribution of $z\sim 1$ early-type galaxies with the
$\mvir$-$\dens$ relation for nearby galaxies.  Because $\dens$ does
not strongly depend on $\mvir$, evolutionary trends are readily
visible; $z\sim 1$ early-type galaxies are $\sim$4 times more dense
than their local counterparts.  The apparent change in slope can
possibly be explained by selection effects, completely analogous to
our conclusion that this is the case with the $\mvir$-$\reff$
relation.  Note that compared to the increase in projected density,
the increase in physical density will be even larger.

Up until recently, early-type galaxies were thought to evolve more or
less passively.  This appears to be an over-simplification and may
apply more to their stellar populations than to their structural
properties.  In the following section we discuss possible explanations
in the context of theoretical predictions and the comparison with
results from studies with different observational strategies.

\section{DISCUSSION}\label{dis}

\subsection{Comparison with Photometric Size-Evolution Measurements}\label{dis:1}

The main goal of this paper is to use dynamical measurements to
investigate whether early-type galaxies were smaller and denser in the
past.  Previous work has shown that the stellar mass surface density
is higher, but there are a number of issues with such studies as they
rely on stellar population models and they ignore possible changes in
the underlying dark matter profile.

\begin{figure}[t]
\epsscale{1.2}
\plotone{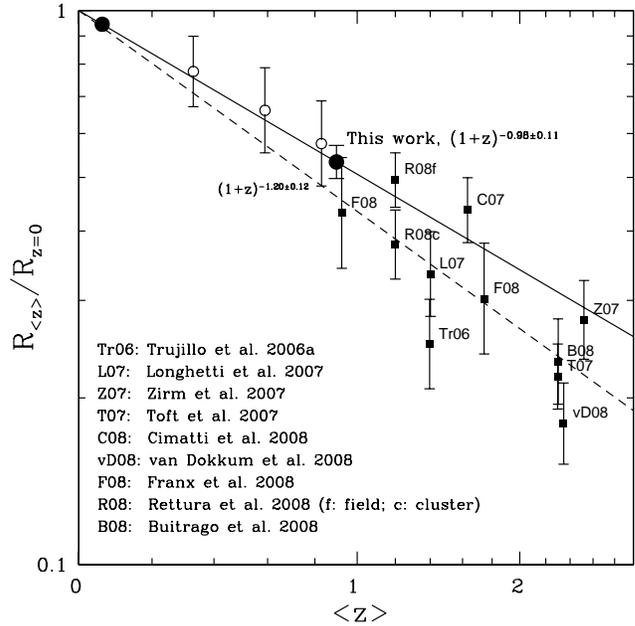}
\caption{Size evolution with redshift as derived in this paper with
  dynamically determined masses (\textit{large filled circles})
  compared with previous results based on photometric masses
  (\textit{small filled circles}).  The solid line connects our
  samples at $z\sim 0.06$ and $z\sim 1$, the dashed line is a linear
  least-squares fit to the small filled data points.  The open circles
  are samples of cluster galaxies with photometrically measured masses
  and serve as an illustration that size evolution shows a continuous
  trend between $z=2.5$ and the present.  The broad agreement in size
  evolution as derived from galaxies with dynamically and
  photometrically determines masses reinforces the conclusions of
  previous, photometric studies whose results were potentially
  mitigated by considerable systematic effects that do not affect our
  analysis. }
\label{lit_z}
\end{figure}

In Fig.~\ref{lit_z} we compare the size-evolution results presented in
the previous section with size-evolution results for early-type
galaxies based on photometric mass estimates.  For all the literature
samples we take the mean redshift and the mean stellar mass
(normalized to the Kroupa IMF), and compute the mean offset from the
local mass-size relation from \citet{shen03}.  We include four
intermediate redshift cluster galaxy samples with photometrically
measured masses and sizes from WFPC2 or ACS imaging.  The data are
described by \citet{holden07} and the sizes are measured as described
in this paper (\S~\ref{distant}).  These four clusters are \clll at
$z=0.33$, \msl at $z=0.59$, and \ms and \cll, both at $z=0.83$.  Note
that we also include \ms in the present study with dynamical mass
measurements.  The agreement between the independent measurements
confirms that at least out to $z\sim 1$ dynamical and photometric mass
estimates based on optical colors and spectral energy distributions
agree within the statistical errors as was previously shown by
\citet{vanderwel06b}, \citet{rettura06}, and \citet{holden06}.

The literature samples have all been selected in different ways, and
so a direct comparison with our work may not be straightforward. Not
all samples are morphologically selected; many are selected by their
spectral or photometric properties. In the local universe there is
substantial overlap between samples of early-type galaxies that are
selected by different criteria; therefore, it is a reasonable
assumption to suppose this to also be the case at high redshift, where
different indicators (low star formation rates, red colors, smooth
visual appearance) also reflect a common nature. Recently, several
studies have shown hints that this is indeed the case
\citep[e.g.,][]{vandokkum08b, kriek08b}, but these issues need to be
further addressed in the future.

Even with this cautionary proviso, the broad agreement between the
results presented in this paper and the photometric results at higher
redshifts is striking. All studies included in Fig.~\ref{lit_z} are
consistent with significant size evolution of several factors between
$z\sim 1-2$ and the present for galaxies with a given mass.  A linear
fit in log-log space to our two data points at $z\sim 0.06$ and $z\sim
1$ gives $\reff(z)\propto (1+z)^{-0.98\pm0.11}$.  With a linear fit to
the photometric data the inferred rate of evolution is
$\reff(z)\propto (1+z)^{-1.20\pm0.12}$, where the error is obtained
via a bootstrap/Monte Carlo simulation.

The broad agreement of our measurement of the size evolution of
early-type galaxies with the photometric studies is encouraging and
alleviates concerns about serious systematic effects that potentially
could have compromised previous work.  Most notably, uncertainties in
the photometric mass estimates used in all other previous work appear
to have a limited impact, at least compared to factors of $\gtrsim$5
which would mimic the strong, observed size and density evolution.
Uncertainties in photometric mass estimates on the level of a factor
of $\sim$2 due to differences among the various stellar population
models \citep[e.g.,][]{bruzual03, maraston05} remain an issue, but to
invoke, for example, an unconventional stellar IMF as an alternative
to radically different structural properties of high-redshift
early-type galaxies is no longer necessary.

Other systematic uncertainties cannot explain the observed evolution
either.  In our size measurements, systematic effects have been taken
into account (see Secs.~\ref{distant:profile} and \ref{sim}).  We are
confident, for example, that we would detect low-surface brightness
envelopes around distant galaxies.  Furthermore, we know that only a
minority of morphologically selected $z\sim 1$ early-type galaxies
($\sim$10\%) show signs of nuclear activity
\citep[e.g.,][]{rodighiero07, vanderwel07a} such that it is unlikely
that central point sources affect our size measurements.  This is also
clear from the fact that the residuals of our $R^{1/4}$ profile fits
generally do not show central point sources and that none of the deep
spectra used to measure dispersions show evidence for nuclear
activity.  Furthermore, the good correspondence between the rest-frame
wavelength of the imaging data sets used at different redshifts assures
us that morphological $K$-corrections do not play a significant role.

Despite the broad consistency between our results and those previously
published, the agreement is not perfect.  There is a marginal
inconsistency at the $1.5~\sigma$ level between the size-evolution
measurement from kinematic data and the size-evolution measurement
from photometric data shown in Fig.~\ref{lit_z}.  This could point to
the presence of some systematic effects within the $z>1.5$ results.
Alternatively, the different studies sample galaxies with a wide range
in masses, and therefore mass-dependent size evolution could lead to
apparent discrepancies among the samples.  This is explored in the
following section.

Our robust results strengthen the results from previous studies.  We
conclude that early-type galaxies at $z=1$ are $\sim 2$ times smaller
than local early types with the same mass, and that at $z=2-2.5$ this
size difference is likely increased to a factor of $\sim$4, as
previously observed by \citet{zirm07}, \citet{toft07},
\citet{vandokkum08b}, and \citet{buitrago08}.

\subsection{Comparison with Model Predictions}\label{dis:2}

The fact that we see considerable evolution in galaxy size with
redshift is not surprising from a theoretical perspective.  Most
semianalytic models of galaxy formation in a $\Lambda$CDM universe
predict substantial size evolution over the past several billion
years.  A comparison between the observed and model-predicted amount
of size evolution will help to identify the mechanism(s) that are
responsible.  In Fig.~\ref{M_dens} we directly compare the observed
evolution in surface density with the predictions from the
semianalytic work by \citet{khochfar06b}.  For galaxies with a given
mass the model significantly under-predicts the evolution in size and
density, except, perhaps, for the most massive galaxies.  In our data
set we see no indication that the magnitude of size and density
evolution increases with galaxy mass, as predicted by the models.  In
fact, the most massive galaxies in our sample are precisely the only
ones that are not different from local massive galaxies.  Note,
however, that statistically speaking the evidence for mass-dependent
evolution is weak (see \S~\ref{results:mrevol}).  Moreover, the most
massive galaxies in our distant sample are a special subset, brightest
cluster galaxies.  Such galaxies have been shown to have properties
that deviate from those of other massive galaxies \citep[see,
e.g.,][]{vonderlinden07, bernardi07b}.

\begin{figure}[t]
\epsscale{1.2}
\plotone{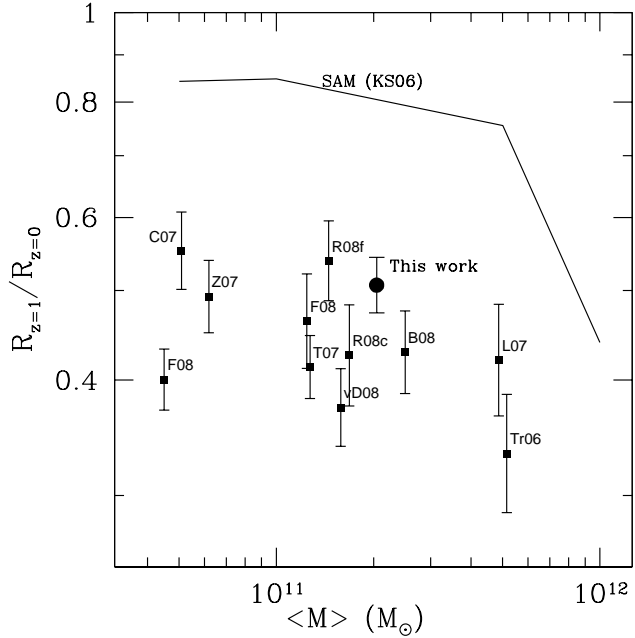}
\caption{Size evolution per unit redshift vs. mean galaxy mass of our
  sample (\textit{large circle}) and samples taken from the literature
  (small squares; see Fig.~\ref{lit_z} for references).  Samples
  consisting of high-mass galaxies show somewhat stronger size
  evolution than samples consisting of low-mass galaxies, which is
  qualitatively, but not quantitatively, consistent with the
  predictions from the semianalytic model from \citet{khochfar06b}
  (\textit{solid line}).  This conclusion should be considered highly
  tentative, however, as this interpretation is hampered by systematic
  uncertainties and small sample sizes. }
\label{lit_M}
\end{figure}

By including the $z=1.5-2.5$ photometric samples discussed in
\S~\ref{dis:1} we can place further constraints on the models.  In
Fig.~\ref{lit_M} we compare the observed size evolution of the
available samples, normalized to $z=1$, with the model predictions
from \citet{khochfar06b}.  Representing the model predictions by a
single line is justified by the fact that the predicted evolution of
$\log(\reff)$ with $\log(1+z)$ is very close to linear.  Again, the
observed size evolution is stronger than that predicted by the model.

It is interesting to note that, in qualitative agreement with the
model prediction, we see a hint that size evolution depends on mass in
the compilation presented in Fig.~\ref{lit_M}.  The samples containing
on average the lowest-mass galaxies display marginally less evolution.
It has to be kept in mind, however, that small sample sizes and
systematic effects are more important for determining second-order
effects such as mass dependence (see \S~\ref{results:mrevol}).  A
clue that systematic uncertainties may play a role is the remaining
difference between the kinematic and photometric samples.
Alternatively, it may signify non-linear evolution of $\log(\reff)$
with $\log(1+z)$.

\subsection{Size Evolution of Individual Galaxies}\label{dis:3}

It appears that the observed size evolution of a factor of $\sim$2
between $z=1$ and the present for early-type galaxies with masses
$\sim10^{11}~\msol$ is similar to the predicted evolution for
early-type galaxies that are an order of magnitude more massive (see
Figs.~\ref{M_dens} and \ref{lit_M}).  This suggests that the mechanism
responsible for increasing the average size of early-type galaxies
with time may be well understood, but that it is not implemented
correctly in the current semianalytic model from \citet{khochfar06b}.
The process of size evolution may occur at different times and under
different circumstances than is now assumed.  This may be related to
the late assembly of very massive galaxies in models of this kind
\citep[see also, e.g.,][]{delucia06}, a prediction that is challenged
by various observations \citep[e.g.,][]{cimatti06, scarlata07,
  cool08}.

It is beyond the scope of this paper to fully discuss these possible
discrepancies.  Instead we will explore the question whether the
proposed physical processes responsible for size evolution are
consistent with the observed trends.  In the semi-analytic models it
is assumed (and this is confirmed by numerical simulations) that
mergers drive size evolution.  The gas content of merging galaxies
largely determines the relative size of the merger remnant compared to
its ancestors.  Because gas fractions were higher in the past,
galaxies that form early will be smaller than galaxies that form late.
In the framework of cosmological simulations this means that galaxies
at high redshift will be smaller because they were formed through
gas-rich mergers and that those merger remnants can grow over time
through subsequent mergers with other galaxies that are progressively
more devoid of gas.

The question is whether the observed size evolution is dominated by
size evolution of individual galaxies or simply by the addition of
larger galaxies over time.  At $z=1$ only about 30--50\% of the
present-day early-type galaxy evolution had formed \citep{bell04b,
  brown07, scarlata07, faber07}.  If we assume that these galaxies
will make up the 30--50\% most dense early-type galaxies in the
present-day universe, then the scatter in the local $\mvir$-$\reff$
relation implies that the sample-averaged size increases by a factor
of 1.3--1.4 between $z=1$ and the present.  Such evolution is thus
expected in the absence of size evolution of individual galaxies and
this is less than the observed evolution of a factor of 2.  To explain
the observed evolution by growth of the early-type galaxy population
without changes in the sizes of individual galaxies, the number
density of early-type galaxies is required to increase by an order of
magnitude between $z=1$ and the present.  Such strong evolution is
clearly ruled out by the above-mentioned determinations of the number
density of red galaxies at $z\sim 1$.

Similarly, at $z=2$ only $\sim$10\% of the galaxies with masses
$\gtrsim~10^{11}~\msol$ had been assembled \citep{kriek08b}; if those
galaxies, evolve into the 10\% most dense present-day early-type
galaxies then an increase in average size by a factor of $\sim$2 can
be accounted for, less than the observed amount of evolution.  These
arguments are in agreement with the conclusions from
\citet{cimatti08}, who show that local galaxies with the same sizes
and masses as galaxies in the $z=1-2$ samples are so rare in the local
universe that it can be confidently ruled out that their structure
remains unchanged up until the present day.  Note, however, that these
arguments may be affected by the aforementioned biases in the SDSS
(\S~\ref{nearby:sigrad}).


We conclude that size evolution due to the addition of larger galaxies
over time contributes at most half of the observed evolution in the
$\mvir$-$\reff$ relation.  The remainder must be due to size evolution
of individual galaxies.  Numerical simulations have demonstrated that
when early-type galaxies accrete neighbors without significant
dissipational processes $\sige$ does not change by much and that, to
first order, $\reff$ increases linearly with mass.  This does not
depend strongly on the mass of the accreted object, i.e., the mass
ratio of the merger \citep{boylan05, robertson06, boylan06}.

Simulations in a cosmological context show that an increase in size by
a factor of 2 between $z\sim 1$ and the present is certainly possible
\citep{naab07}.  The strong observed size evolution thus argues in
favor of a scenario in which significant mass from low-mass companions
is accreted onto existing early-type galaxies over the past $\sim$7
Gyr, which also explains the broad tidal features that are frequently
observed around early-type galaxies \citep{vandokkum05}.  As shown by
\citet{feldmann08} such features are not necessarily, and are even
quite unlikely to be, the result of major merger events and are most
likely due to the accretion of low-mass, gas-poor satellites.

We note that the size evolution of individual galaxies and the
evolution of the sample average are inseparable because galaxies
evolve in mass as well as in size.  Nonetheless, it is important to
distinguish this complex scenario from the simple picture in which
early-type galaxies that form at different redshifts have different
sizes but do not structurally evolve at later times.  The strong
observed size evolution clearly rules out the latter, indicating that
the build-up of the early-type galaxy population is a complex and
ongoing process.

Finally, it is remarkable that the change in the sizes of early-type
galaxies is consistent with and differs by less than 15\% from the
change in the scale factor of the universe, $1+z$.  Within the
standard cold dark matter scenario this is likely a coincidence since
dissipational, strongly non-linear processes that are decoupled from
cosmic expansion dominate at the kiloparsec scale of forming galaxies.
Nonetheless, we cannot exclude the possibility that there is an
underlying, fundamental reason that galaxies are scale-invariant with
respect to a co-moving coordinate system.  In an alternative
description of dark matter, i.e., Bose-Einstein condensed, ultra-light
particles with a $\sim$10 kpc-sized wave function \citep[fuzzy dark
matter or FDM,][]{sin94,hu00}, sizes of halos and their occupying
galaxies possibly follow the cosmic expansion rate \citep{lee08}.

\section{SUMMARY}\label{sum}

In \S~\ref{nearby} we construct a large sample of nearby
($0.04<z<0.08$) early-type galaxies extracted from the SDSS (DR6).  We
use the pipeline velocity dispersion measurements and obtain our own
size measurements in order to construct the local dynamical mass-size
relation (\S~\ref{nearby:mr}).  In addition, we construct a sample of
50 morphologically selected early-type galaxies in the redshift range
$0.8<z<1.2$ with measured velocity dispersions (\S~\ref{distant}).
Sizes are determined from ACS imaging in the same manner as for the
galaxies in the nearby sample, and systematic effects are quantified
through simulations (\S~\ref{sim}).  The distant sample contains
galaxies in the mass range $3\times 10^{10}~\msol < M \lesssim
10^{12}~\msol$, with a typical mass of $2\times 10^{11}~\msol$.

The main result is that the $\sige$-$\reff$ distributions of the
nearby and distant samples are significantly different, even after we
correct for the incompleteness of the distant sample at low masses
(\S~\ref{results}).  The implied size evolution is $\reff \propto
(1+z)^{-0.98 \pm 0.11}$, or a factor of $1.97 \pm 0.15$ between $z=1$
and the present.  Similarly, the projected surface densities of the
distant early-type galaxies are a factor of $\sim$4 higher than those
of their local counterparts.  The stellar populations of early-type
galaxies that already existed at $z=1$ may, for the most part, be
passively evolving over the past 7--8 Gyr, however, their structural
properties undergo substantial changes over that period.

Our results are in broad agreement (see \S~\ref{dis:1}) with
previously published size-evolution measurements that are based on
samples without dynamical mass measurements and, in some cases,
without spectroscopic redshifts, high-resolution \textit{HST} imaging,
and/or consistently determined sizes.  We therefore conclude that
systematic effects, most notably those in the mass estimates, which
potentially could have hampered previous studies are small relative to
the observed amount of evolution.

The observed size evolution is in qualitative agreement with
predictions from recent semianalytic models.  However, the predicted
evolution is much slower than the observed evolution.  The observed
size evolution of early-type galaxies can be understood within the
context of the cold dark matter scenario in which galaxies that form
late have larger sizes than galaxies that form early, due to lower gas
fractions at late times, and the growth of individual galaxies through
the mostly dissipationless accretion of satellites at later
evolutionary stages.

\acknowledgements{We thank the referee for his helpful
  review. A.~v.~d.~W. would like to thank Jenny Graves for sharing her
  SDSS catalog of early-type galaxies, Eric Bell, Hans-Walter Rix, and
  Pieter van Dokkum for interesting discussions, and Sadegh Khochfar
  for providing his model predictions.  Support from NASA grant
  NAG5-7697 is also gratefully acknowledged.  The authors wish to
  recognize and acknowledge the very significant cultural role and
  reverence that the summit of Mauna Kea has always had within the
  indigenous Hawaiian community.  We are most fortunate to have the
  opportunity to conduct observations from this mountain.  Funding for
  the SDSS and SDSS-II has been provided by the Alfred P. Sloan
  Foundation, the Participating Institutions, the National Science
  Foundation, the U.S. Department of Energy, the National Aeronautics
  and Space Administration, the Japanese Monbukagakusho, the Max
  Planck Society, and the Higher Education Funding Council for
  England. The SDSS Web Site is http://www.sdss.org/.  The SDSS is
  managed by the Astrophysical Research Consortium for the
  Participating Institutions. The Participating Institutions are the
  American Museum of Natural History, Astrophysical Institute Potsdam,
  University of Basel, University of Cambridge, Case Western Reserve
  University, University of Chicago, Drexel University, Fermilab, the
  Institute for Advanced Study, the Japan Participation Group, Johns
  Hopkins University, the Joint Institute for Nuclear Astrophysics,
  the Kavli Institute for Particle Astrophysics and Cosmology, the
  Korean Scientist Group, the Chinese Academy of Sciences (LAMOST),
  Los Alamos National Laboratory, the Max-Planck-Institute for
  Astronomy (MPIA), the Max-Planck-Institute for Astrophysics (MPA),
  New Mexico State University, Ohio State University, University of
  Pittsburgh, University of Portsmouth, Princeton University, the
  United States Naval Observatory, and the University of Washington. }


\begin{deluxetable}{lccc}
\tabletypesize{\scriptsize}
\tablecolumns{4}
\tablewidth{0pt}
\tablenum{1}
\tablecaption{Velocity Dispersions and Sizes of the Distant Sample}
\tablehead {
\colhead{ID} &
\colhead{$z$} &
\colhead{$\sige$} &
\colhead{$\reff$}}
\startdata
MS 1054-1649 & 0.831 & $243\pm28$ &  4.91 \\
MS 1054-2409 & 0.831 & $287\pm33$ &  3.30 \\
MS 1054-3058 & 0.831 & $303\pm33$ & 10.20 \\
MS 1054-3768 & 0.831 & $222\pm24$ &  3.28 \\
MS 1054-3910 & 0.831 & $295\pm42$ &  1.80 \\
MS 1054-4345 & 0.831 & $336\pm34$ &  4.35 \\
MS 1054-4520 & 0.831 & $322\pm30$ & 15.20 \\
MS 1054-4705 & 0.831 & $253\pm36$ &  8.84 \\
MS 1054-4926 & 0.831 & $310\pm38$ &  2.04 \\
MS 1054-5280 & 0.831 & $259\pm31$ &  3.68 \\
MS 1054-5298 & 0.831 & $284\pm39$ &  3.54 \\
MS 1054-5347 & 0.831 & $254\pm24$ &  2.94 \\
MS 1054-5450 & 0.831 & $234\pm26$ &  8.16 \\
MS 1054-5529 & 0.831 & $182\pm23$ &  3.24 \\
MS 1054-5577 & 0.831 & $305\pm40$ &  2.67 \\
MS 1054-5666 & 0.831 & $286\pm23$ &  4.99 \\
MS 1054-5756 & 0.831 & $232\pm27$ &  3.98 \\
MS 1054-6036 & 0.831 & $254\pm22$ &  2.93 \\
MS 1054-6301 & 0.831 & $249\pm24$ &  3.55 \\
MS 1054-6688 & 0.831 & $274\pm37$ &  2.93 \\
   HDFN-206 & 0.936 & $199\pm18$ &  1.11 \\
   HDFN-237 & 0.851 & $280\pm21$ &  1.80 \\
   HDFN-256 & 0.974 & $306\pm14$ &  3.06 \\
   HDFN-635 & 0.820 & $201\pm17$ &  2.29 \\
   HDFN-681 & 0.842 & $341\pm30$ &  1.43 \\
   HDFN-761 & 1.013 & $374\pm39$ &  3.77 \\
   HDFN-811 & 0.848 & $216\pm14$ &  1.32 \\
   HDFN-933 & 0.847 & $305\pm37$ &  1.86 \\
   HDFN-951 & 0.854 & $235\pm17$ &  2.15 \\
  HDFN-1236 & 0.850 & $217\pm12$ &  1.98 \\
  HDFN-1286 & 0.846 & $247\pm17$ &  3.66 \\
  HDFN-1287 & 0.846 & $342\pm23$ &  3.94 \\
  HDFN-1328 & 0.845 & $250\pm34$ &  1.82 \\
  HDFN-1543 & 0.849 & $280\pm13$ &  1.52 \\
  HDFN-1559 & 0.943 & $178\pm12$ &  1.66 \\
  HDFN-1633 & 0.841 & $330\pm13$ &  1.96 \\
  HDFN-1706 & 0.913 & $215\pm12$ &  2.23 \\
  HDFN-1709 & 0.842 & $218\pm11$ &  1.16 \\
     CDFS-1 & 1.089 & $231\pm16$ &  2.83 \\
     CDFS-2 & 0.964 & $200\pm10$ &  2.30 \\
     CDFS-3 & 1.044 & $300\pm32$ &  1.00 \\
     CDFS-4 & 0.964 & $336\pm19$ &  6.84 \\
     CDFS-7 & 1.135 & $232\pm20$ &  5.77 \\
    CDFS-12 & 1.123 & $262\pm21$ &  0.94 \\
    CDFS-13 & 0.980 & $247\pm11$ &  2.20 \\
    CDFS-14 & 0.984 & $197\pm23$ &  2.80 \\
    CDFS-18 & 1.096 & $324\pm36$ &  3.97 \\
    CDFS-20 & 1.022 & $199\pm16$ &  2.24 \\
    CDFS-25 & 0.967 & $258\pm19$ &  0.86 \\
    CDFS-29 & 1.128 & $221\pm18$ &  1.59 \\
\enddata

\tablecomments { The IDs and velocity dispersions are taken from
  \citet{wuyts04}, \citet{treu05b}, and \citet{vanderwel05} for \ms,
  HDFN, and CDFS, respectively.  The dispersions are all
  aperture-corrected according to eq.~\ref{eq:apcor}.  The sizes are
  determined as described in \S~\ref{distant:sigrad}, and a standard
  error of 14\% is adopted for all galaxies, which is based on the
  simulations described in \S~\ref{sim}.  }

\label{tab}
\end{deluxetable}

\end{document}